\documentclass[12pt,a4paper]{article}
\usepackage{graphicx}% Include figure files

\begin{document}
\textwidth=135mm
 \textheight=200mm
\begin{center}
{\bfseries Looking for antineutrino flux from $^{40}$K with large liquid scintillator detector 
\footnote{{\small Report at the International Workshop on Prospects of Particle Physics: «Neutrino Physics and Astrophysics», Valday, January 27 -
February 2, 2014.}}}
\vskip 5mm
\underline{V. V. Sinev}$^a$, L. B. Bezrukov$^a$, E. A. Litvinovich$^{b,c}$, I. N. Machulin$^{b,c}$, M. D. Skorokhvatov$^{b,c}$ and S. V. Sukhotin$^b$
\vskip 5mm
{\small {\it $^a$ Institute for Nuclear Researches Russian Academy of Sciences, Moscow, Russia}} \\
{\small {\it $^b$ National Research Center Kurchatov Institute, Moscow, Russia}}\\
{\small {\it $^c$ National Research Nuclear University MEPhI (Moscow Engineering Physics Institute) }}
\\
\end{center}
\vskip 5mm

\centerline{\bf Abstract}
We regard a possibility of detecting the antineutrino flux produced by the $^{40}$K located inside the Earth. 
Thermal flux of the Earth could be better understood by observing such a flux.
Lower and upper limits on the $^{40}$K antineutrino flux are presented.
\vskip 10mm

\section{Introduction}

Measured Earth thermal flux is estimated to be 30-40 TW [1]. 
This is approximately 60-80 mW/m$^2$ on average. 
We do not know exactly all heat sources producing a flux like this. 
According to the element abundances in Bulk Silicate Earth model, 
the amount of radioactive isotopes can explain only about 20 TW of the total thermal flux. 
It is produced by decays of $^{238}$U, $^{232}$Th and $^{40}$K [2, 3]. 
Some exotic thermal sources like natural nuclear reactor placed in the Earth core are also considered [4].

Modern neutrino registration methods can help us to measure the amount of radioactivity in the Earth. 
All radioactive elements and georeactor emit antineutrinos which pass through the Earth and go into space without interaction. 
A large scintillating detector near the surface can detect these antineutrinos 
and obtain the total antineutrino flux which is proportional to the total mass of radioactive constituents of the Earth.

Antineutrino can be detected through the inverse beta-decay reaction (IBD) 
that produces pair events in the detector clearly distinguished from backgrounds. 
However, it has a relatively high threshold of 1.8 MeV that discards $^{40}$K antineutrinos having maximal energy 1.3 MeV. 
$^{40}$K antineutrinos can be detected only through the reaction of elastic scattering on electrons, cross section of which is two orders lower than IBD one.

Geoneutrino flux from $^{238}$U and $^{232}$Th was detected recently by two scintillation detectors of large volume: KamLAND (1000 t) [5] and BOREXINO (300 t) [6]. 
The measured geoneutrino flux does not contradict the minimal abundances of $^{238}$U and $^{232}$Th that follow from Bulk Silicate Earth model,
but also does not discard other theories with larger abundances [7, 8]. 
The experimental uncertainty is about 25\%. 
The georeactor power is limited by 4.5 TW from these measurements.

In [9] they regard geoneutrino flux produced by $^{40}$K. 
If potassium abundance achieves the value of 3.76\% (from [8]), 
the flux becomes comparable to the $^{7}$Be flux from the Sun: 7.8$\times10^8$ cm$^{-2}$ s$^{-1}$ contra 4.8$\times10^9$ cm$^{-2}$ s$^{-1}$ of $^7$Be flux. 

Potassium abundance in the Earth varies in a number of works from 0.024\% [3] up to 3.76\% [8]. 
We analyzed the influence of potassium abundances in Earth layers on $^{40}$K antineutrino flux and estimated upper and lower limits of potassium flux. 
In contrast to [3], we placed most of potassium in the core and observable abundance in the crust. 
We did calculations for the same detector volume and scintillator as the one used in Borexino.

\section{$^{40}$K antineutrino flux}

$^{40}$K decay scheme is shown in fig. 1 [10, 11]. 
The main transition with probability 89.25\% goes to the ground state of $^{40}$Ca emitting a beta-particle and antineutrino with end-point energy 1.311 MeV. 
In 10.55\% of events there is K-capture from the excited level of $^{40}$Ar with the emission of monoenergetic 44 keV neutrino, 
the nucleus then emitting a photon with energy 1.46 MeV returning $^{40}$Ar to the ground state. 
In 0.2\% K-capture leads to a decay to the ground state of $^{40}$Ar with the emission of 1.5 MeV monoenergetic neutrino.

\begin{figure}[ht] 
\vspace{10pt}
\label{fig:f1} 
\centering
\includegraphics[scale=1.2]{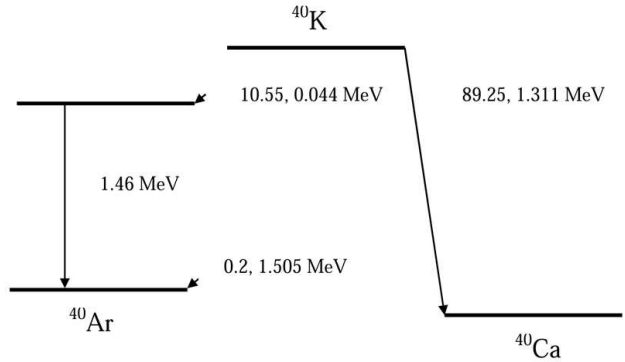}
\caption{Decay scheme of $^{40}$K.}
\end{figure}

\begin{figure}[ht] 
\label{fig:f2} 
\centering
\includegraphics[scale=1.2]{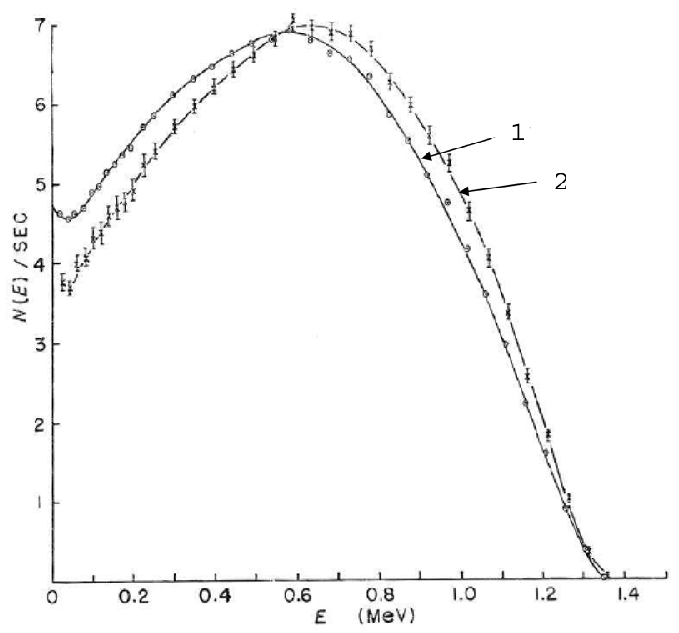}
\caption{Beta-spectra for $^{40}$K: 1 $-$ experimental measurement including background; 
2 $-$ beta-spectrum after subtracting background and corrected by the detector response function.}
\end{figure}

We calculated a $^{40}$K antineutrino spectrum corresponding to the beta-spectrum shown in fig. 2 [3, 12]. 
Our beta-spectrum differs from the experimental one, but it still can be used to estimate the effect of antineutrinos on a detector. 
In the first approximation one does not need to use weak magnetism corrections 
for antineutrino spectrum, since even though they are large enough for beta-particles, they are small for antineutrinos. 
That is why we do not use any corrections for antineutrino spectrum shown in fig. 3.

\begin{figure}[ht] 
\vspace{10pt}
\label{fig:f3} 
\centering
\includegraphics[scale=1.2]{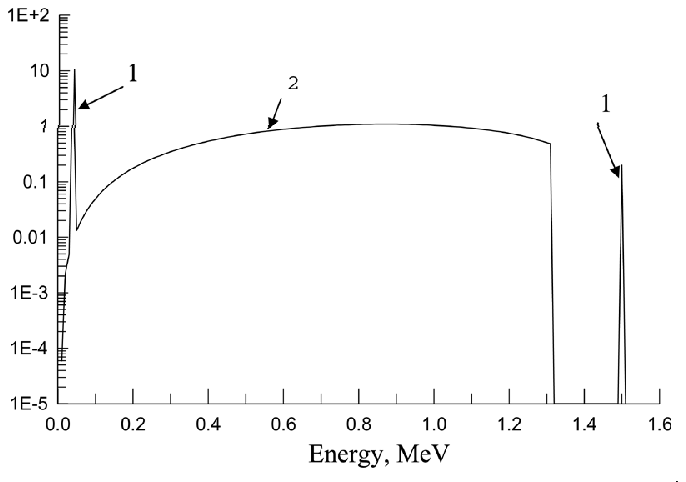}
\caption{Neutrino spectrum from $^{40}$K according to the decay scheme shown in fig.1 
(1 $-$ neutrinos, 2 $-$ antineutrinos); Y-axis shows probability of producing (anti)neutrino per MeV per one decay of $^{40}$K.}
\end{figure}

\section{Neutrino detector}

We consider liquid organic scintillator as the detector target. 
In the BOREXINO detector the scintillator used is based on Pseudocumene (PC), and in KamLAND detector it is based on mineral oil. 
However, in some modern detectors the use of scintillator based on Linear alkyl benzene (LAB) is proposed. 
In table 1 we show numbers of Carbon, Hydrogen and electrons containing in 1000 t of LAB and PC.

\begin{table}[ht]
\caption{Abundance of H, C and electrons in 1000 t linear alkyl benzene and pseudocumene.}
\centering
\label{table:1}
\begin{tabular}{l|c|c}
\hline
1000 t & LAB &  PC \\
\hline
Formula & C$_{18}$H$_{30}$ & C$_{9}$H$_{12}$ \\
H & $7.465\times10^{31}$ & $6.013\times10^{31}$ \\
C & $4.479\times10^{31}$ & $4.510\times10^{31}$ \\
Electrons & $3.434\times10^{32}$ & $3.307\times10^{32}$ \\
\hline
\end{tabular}\\[2pt]
\end{table} 

\begin{table}[ht]
\caption{Potassium abundance in Earth`s layers (\% by weight). 
To change to g/g units one should use coefficient $10^{-2}$.}
\label{table:2}
\begin{tabular}{l|c|c|c}
\hline
 & Min. abund. & Max. abund. & Max. abund. \\
 & \% &  (Model 1), \% &  (Model 2), \% \\
\hline
Crust & 2.1 & 2.1 & 2.1 \\
Upper mantle & 2.1 & 2.1 & 3.0 \\
Lower mantle & 0.0 & 0.0 & 3.5 \\
Outer core & 0.0 & 10.0 & 4.5 \\
Inner core & 0.0 & 10.0 & 6.0 \\
\hline
Oceans & 0.042 & 0.042 & 0.042 \\
Sediments & 0.2 & 0.2 & 0.2 \\
\hline
Total & 0.36 & 3.74 & 3.74 \\
\hline
\hline
\end{tabular}\\[2pt]
\end{table} 

\begin{table*}[ht]
\caption{Antineutrino and neutrino fluxes and their effects in 100 t of pseudocumene.}
\label{table:3}
\vspace{10pt}
\begin{tabular}{l|c|c|c|c}
\hline
  & Neutrino & Antineutrino & Neutrino & Antineutrino \\
\hline
Flux, cm$^{-2}$ s$^{-1}$ & $1.70\times10^{6}$ & $7.58\times10^{8}$ & $3.05\times10^{5}$ & $1.36\times10^{8}$ \\
\hline
Rate, d$^{-1}$ & 0.06 & 4.04 (Mod.1) & 0.015 & 1.01 \\
  &   & 4.54 (Mod.2) &   &  \\
\hline
\hline
\end{tabular}\\[2pt]
\end{table*} 
\vspace{30pt}

Antineutrinos from $^{40}$K are registered through the reaction of elastic scattering of antineutrinos on target electrons 
\begin{equation}
\bar{\nu_{e}} + e^{-} \rightarrow \ \bar{\nu_{e}^{\prime }} + e^{-\prime }. 
\end{equation}
Cross section of reaction (1) is written as 
\begin{eqnarray}
\frac{d\sigma^W}{dT}=\frac{g^{2}_{F}}{\pi} m \times \Bigl[(1+2x^2)^2\left(1-\frac{T}{E}\right)^2 + \nonumber \\
4x^2-2x^2(1+2x^2)\frac{mT}{E^2}\Bigr], \quad 
\end{eqnarray}
where $E$ and $T$ are antineutrino energy and electron recoil energy respectively, $g^{2}_{F}\frac{m}{\pi}=4.308\times10^{-45}$ cm$^2$, $x^2=\sin^2\theta_W=0.232$.

\section{Earth models testing}

To do calculations we have chosen the Earth model of concentric spheres according to seismic data. 
From the surface down to Mohorovicic`s boarder is the crust which is divided into upper, middle and lower parts. 
We took data on the depth of these parts from [13], where the data are presented 
divided into bins with altitude and longitude changing in steps of two degrees. 
Then, 660 km further down, there is an upper mantle, which we also accounted as lithosphere. 
At the depth 2900 km lower mantle changes into outer (liquid) core which lasts until 5140 km, then down to the Earth centre continues inner (solid) core.

We placed $^{40}$K in the crust and upper mantle only when calculating lower limit for antineutrino flux. 
We took potassium abundance as 2.1 weight \% for the crust and upper mantle according to [14] (mean value appeared to be 0.3\%). 
When calculating upper limit we also added potassium in the solid and liquid cores in 
a certain amount to achieve 3.7\% abundance for the whole Earth which is in agreement with Hydridic Earth model [7, 8]. 
Data used by us are shown in the table 2. 

In table 3 one can find calculated antineutrino and neutrino fluxes from $^{40}$K 
and expected effects for the target of 100 t pseudocumene with threshold 200 keV. 
Vacuum oscillations were taken into account when doing the calculations while the MSW effect was not. 
In addition, muon and tauon neutrinos appearing in oscillations also participate in scattering. 

Recoil electron spectra for minimal and maximal potassium abundances are shown in fig. 4 in comparison with single events from solar neutrino fluxes and inner backgrounds of the BOREXINO detector[15]. 
Daily counting rate is shown in parentheses. 
The counting rate for the flux from $^7$Be solar neutrino flux is 46 per day and $^{40}$K antineutrino flux rate is from 1 to 4 per day.

\begin{figure}[hbt] 
\label{fig:f4} 
\centering
\includegraphics[scale=1.25]{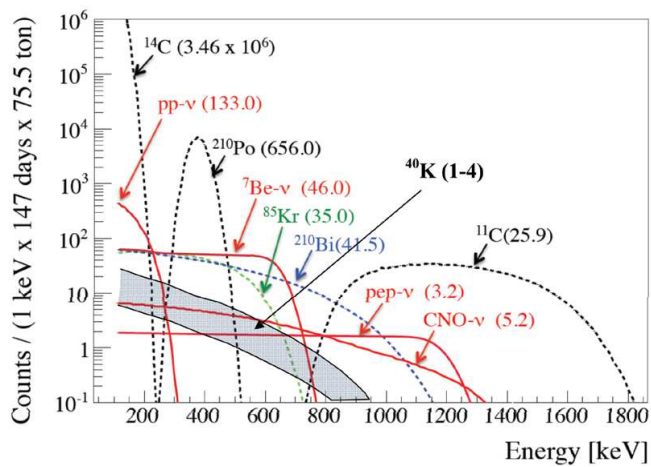}
\caption{Single events spectra in the energy range of BOREXINO per 147 days of data taking in 75.5 t of liquid scintillator [15]. 
The corridor of the possible values of $^{40}$K antineutrino flux caused  single events spectrum is shown.}
\end{figure}

\section{Conclusion}

We present calculations of recoil electron spectra produced by antineutrino flux from isotope $^{40}$K located inside the Earth. 
It appears that the $^{40}$K antineutrino flux is comparable with the neutrino flux produced by $^{7}$Be in solar flux measured with BOREXINO. 
This background was never regarded as significant for solar neutrino detectors. 
The calculations show that it can achieve $\sim$10\% of beryllium neutrinos effect depending on potassium abundance in the Earth. 

The result of our estimation does not contradict the BOREXINO experimental data even in the case of maximal abundance according to [7, 8].

Detector of BOREXINO type can register $^{40}$K antineutrino spectrum and establish the upper limit on potassium abundance in the Earth. 
In the near future the Borexino Collaboration will make an effort to decrease the existing background level. 
This definitely increases the probability of determining $^{40}$K antineutrino flux.
The detection of extra amount of $^{40}$K can help to find an additional heat source for the Earth thermal flux. 

The work was supported by RFBR grants 12-02-12124 and 12-02-92440.

\end{document}